\documentclass{emulateapj}

\usepackage{lineno}

\begin{document}

\renewcommand{\topfraction}{1.0}
\renewcommand{\bottomfraction}{1.0}
\renewcommand{\textfraction}{0.0}

\title{Speckle interferometry of secondary components in nearby visual binaries\altaffilmark{\dag} }

\altaffiltext{\dag}{Based on observations obtained  at Gemini-N. }

\author{Andrei Tokovinin}
\affil{Cerro Tololo Inter-American Observatory, Casilla 603, La Serena, Chile}
\email{atokovinin@ctio.noao.edu}
\author{Elliott P. Horch\footnote{Adjunct Astronomer, Lowell Observatory} }
\affil{Department of Physics, Southern Connecticut State University, 501 Crescent Street, New Haven, CT 06515, USA}
\email{horche2@southernct.edu}

\begin{abstract}
Statistical characterization of secondary subsystems in binaries helps
to distinguish  between various scenarios  of multiple-star formation.
The DSSI  speckle instrument  was used at  the Gemini-N  telescope for
several  hours  in  2015  July  to  probe  binarity  of  25  secondary
components  in nearby  solar-type binaries.   Six new  subsystems were
resolved, with meaningful detection  limits for the remaining targets.
The large incidence of  secondary subsystems agrees with other similar
studies.   The  newly  resolved  subsystem  HIP  115417  Ba,Bb  causes
deviations in  the observed motion of  the outer binary  from which an
astrometric orbit of Ba,Bb with a period of 117 years is deduced.
\end{abstract} 

\section{Introduction}
\label{sec:intro}

Formation of multiple  systems is an active research  topic related to
such areas as stellar mass function, disks, and initial conditions for
planet formation. Although large-scale numerical simulations
give reasonable match to  the observed statistics \citep{Bate2012}, we
are  still  far  from  modeling  the  distributions  of  multiple-star
parameters in  a predictive way.  The relevant  physics is identified,
but the correct mix  of processes that define multiple-star population
is yet to be found.

Stochastic  dynamics of small  $N$-body systems  is one  such process.
Recently \citet{RM12}  suggested that wide binaries  are mostly formed
by ejections from smaller $N$-body  aggregates.  It is well known that
in chaotic dynamics the  smallest-mass stars are ejected most readily.
We therefore would expect wide  companions to be preferentially of low
mass, single,  and on eccentric  orbits.  Another physical  process --
rotationally-driven  fragmentation --  makes  the opposite  prediction
\citep{Delgado2004}.   In this  case,  wide binaries  contain a  large
fraction  of the  collapsing cloud's  angular momentum.   Their orbits
should have moderate eccentricity,  the masses of both wide components
should be comparable,  and the incidence of subsystems  in the primary
and secondary components should be similar.

In this paper we focus on multiplicity of secondary components of wide
binaries, so far poorly characterized. The primary components of those
binaries   are  nearby  solar-type   dwarfs;  their   multiplicity  is
constrained by  combination of  methods, from radial  velocities (RVs)
for close pairs to imaging and  wide common-proper-motion (CPM)
binaries, spanning  the full range of  periods \citep{FG67a}.  In
contrast, the fraction of subsystems in the secondary components could
be  estimated  with  a  considerable  uncertainty.  Yet,  this  is  an
important  diagnostic of  the formation  processes.  If   secondary
components were ejected on their wide orbits by dynamical interplay in
compact  and  unstable  nascent   multiple  systems,  they  should  be
preferentially single.  

In recent years, the binarity  of secondary components has been probed
by  several  techniques. Adaptive  optics  (AO)  imaging  was used  at
Gemini-S  \citep{Tok2010}, Palomar 1.5-m  Robo-AO \citep{Law2010,RAO},
and SOAR \citep{Tok2014}; RVs  were monitored by \citet{CHIRON}.  This
project  explores the  binarity of  secondary components  by  means of
speckle interferometry at the  8-m Gemini-N telescope. Compared to the
Robo-AO, a 7-fold  increase in the angular resolution  gives access to
much shorter periods,  exploring a larger part of  the parameter space
and overlapping with the RV surveys.

In Section~2 we describe the instrument, observations, and data
reduction. Newly resolved binaries and detection limits are presented
in Section 3. In Section 4 we analyze one of the newly discovered
subsystems and determine its orbit using archival measurements of the
wide outer binary. Section 5 concludes the paper.

\begin{figure}
\epsscale{1.1}
\plotone{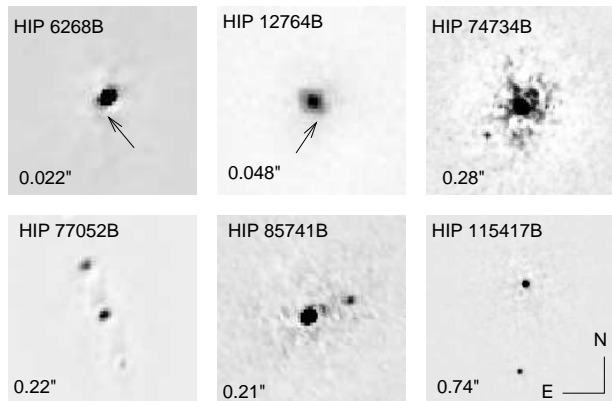}
\caption{Negative images of newly  resolved subsystems at 880\,nm. The
  scale and  intensity stretch are adjusted  individually to highlight
  the companions. The  numbers in the lower left  corner of each image
  are angular separations.  The first two resolutions (HIP 6268B
    and HIP 12764B) are tentative. 
\label{fig:mosaic} }
\end{figure}

\begin{figure*}
\epsscale{1.1}
\plotone{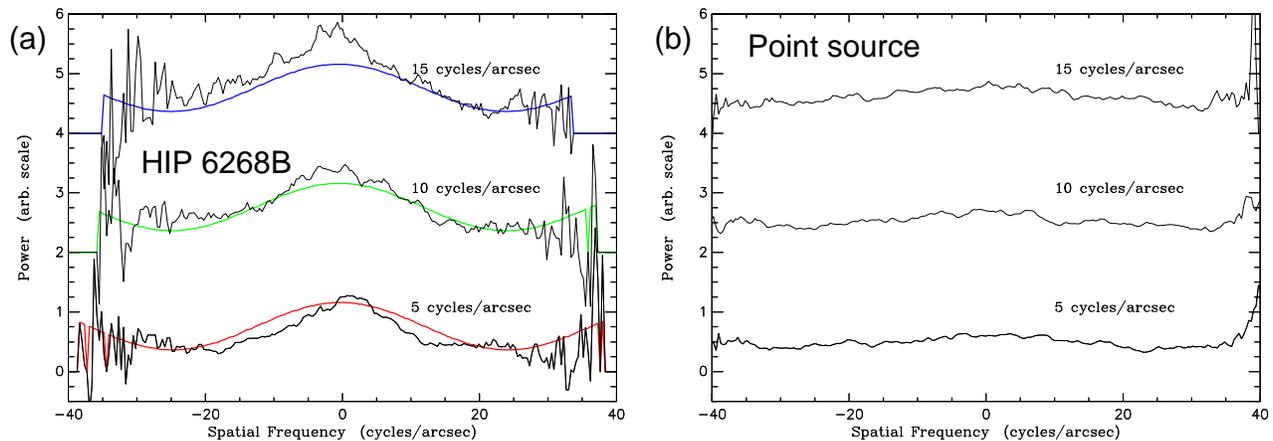}
\caption{Left  panel (a): cuts in  the final fitted
  power  spectrum  perpendicular  to  the  fringe  direction  for  the
  marginal detection  of HIP  6268 Ba,Bb at  880\,nm.  Three  cuts are
  shown, each  of which has  a closest approach  to the origin  of the
  uv plane  of 5, 10,  and 15  cycles per  arcsecond as  shown (i.e.
  this  value is  measured  parallel to  the  fringe direction).   The
  fitted binary model  is plotted with the colored  curves.  A similar
  analysis for a point source, where  no fit is presented, is shown in
  the right panel (b).  This gives a sense of how flat a typical point
  source observation  is in the  frequency domain and  how significant
  the binary signal is. \label{fig:fringes} }
\end{figure*}

\section{Observations and data reduction}
\label{sec:obs}

Observing time  for this work  has been granted through  NOAO (program
15A-0087) for  a total of  7 hours, with  a low Band 3  priority. This
project  was a  ``filler'' for  more challenging  observations,  as it
could be  executed in less than optimum  conditions.  The observations
at Gemini-N were conducted by  E.H., Mark Everett (NOAO), Steve Howell
(NASA-Ames), Johanna  Teske (DTM and Carnegie  Observatories), and Lea
Hirsch (UC Berkeley),  between 2015 July 11 and  19.  During this run,
the sky was clear for about 2/3 of the time, allowing to use 4.7 hours
for this program, part of it through the clouds.

The  target list was  based on  the 67-pc  sample of  solar-type stars
\citep{FG67a}.   Known  binaries  with  separations from  0\farcs5  to
2\farcs8 that  had no  prior high-resolution data  from AO  or speckle
instruments were  selected. Such binaries  fit in the camera  field of
view, allowing  to check  both components for  the existence  of close
subsystems. In  addition, secondary components of  wider binaries with
separations  from 3\arcsec  ~to 10\arcsec  ~were  targeted separately.
The program contained 63 targets; for 39 of those useful data could be
obtained.  In some cases, the brighter component A was pointed instead
of the intended secondary B, either by error or because the extinction
was too  great to successfully  observe the secondary while  the primary
was still visible.

The  Differential Speckle  Survey  Instrument, DSSI,  is described  by
\citet{Horch2009}. Speckle images of the observed star are recorded by
two electron multiplication CCDs simultaneously in two spectral bands,
the light  being divided by a  dichroic. At Gemini-N,  the DSSI worked
with the  filters that transmit central wavelengths  and bandwidths of
692/40 and 880/50  nm; for brevity they are called  here bands $R$ and
$I$.  Previous publications resulting  from the DSSI at Gemini-N, e.g.
\citep{Horch2012,Horch2015a},   contain  additional  details.    In  a
typical observing  sequence, 1000 frames  with a 60-ms exposure  and a
256$\times$256   size    are   recorded   in    both   DSSI   channels
simultaneously. Observations of  single unresolved reference stars are
taken at low  airmass and used for modeling  instrument signatures and
atmospheric  dispersion during  data  reduction.  The  data cubes  are
processed  by the  standard  speckle technique  (calculation of  power
spectrum and autocorrelation) and  by the speckle image reconstruction
delivering     true      images     \citep[see     further     details
  in][]{Horch2009,Horch2015b}.   Figure~\ref{fig:fringes}  illustrates
the resolution of a very close binary HIP~6268B.

The pixel  scale of 11.41\,mas  and the orientation of  both detectors
was  calibrated as  described by  \citet{Horch2012}.  We  observed two
binaries,  HIP~83838 =  HU~1176 and  HIP~104858 =  STT~535,  for which
extremely    accurate    interferometric    orbits    are    available
\citep{Muterspaugh2008,Muterspaugh2010}.  After correcting the optical
distortion in the reflective channel of the DSSI \citep{Horch2011}, we
found  astrometry  in the  two  channels  to  be in  excellent  mutual
agreement at a level of $\sim$2\,mas  for most pairs, in line with the
previous  analysis  of  Gemini-N  speckle data  in  \citep{Horch2012};
larger discrepancies are present only  for pairs near the limit of the
technique. 

Although  the   detectors used  with DSSI  have  a 512$\times$512
pixel  format, the  speckle  frames were  sub-arrays of  256$\times$256
pixels centered  on the target, so  the field of view  of the recorded
images was 2\farcs8$\times$2\farcs8.  This allows detection of companions
with separations  up to  1\farcs45. Some wider  binaries (up  to about
2\arcsec ~separation) were measured  by placing both components in the
field, i.e. centering the sub-array in between the two sources.

\begin{figure}
\epsscale{1.1}
\plotone{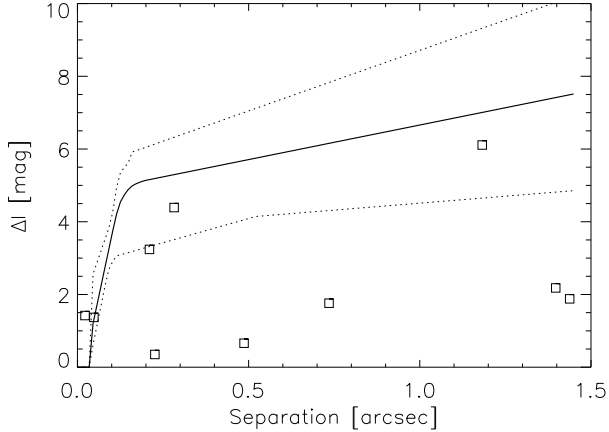}
\caption{Detection limits in the $I$ channel of DSSI. The full line is
  the average detection curve, the dotted lines are the best and worst
  detection limits. Squares denote the actually measured binaries.
\label{fig:det} }
\end{figure}

\section{Results}
\label{sec:res}

\subsection{Measurements}

Binary companions,  either known or newly  discovered, were identified
visually   in  the  reconstructed   images  (Figure~\ref{fig:mosaic}).
Relative  astrometry and  photometry  of binary  stars  is derived  by
approximating the speckle power spectrum, as described in the previous
DSSI publications   \citep{Horch2009,Horch2012}.   The detection
limits are estimated by computing rms fluctuations $\sigma$ in annular
zones of the  reconstructed image and assuming that  a companion above
$5   \sigma$  would   be  detectable,   see   \citep{Horch2011}.  
  Figure~\ref{fig:det}  illustrates the  detection limits  in  the $I$
  channel.

High  contrast  ($\Delta  m$  of   4  to  6  mag)  and  high-SNR  DSSI
observations at Gemini, such as those described here, generally detect
companions  down  to a  separation  of  about  0\farcs1.  For  smaller
separations, the dynamic range  decreases to something below $\Delta m
\sim 1$ mag at the diffraction  limit.  We have looked for evidence of
vibration effects  in our data  by studying the centroid  positions of
bright stars as  a function of time, but have  not found any signature
of  vibration to  the limit  of our  time resolution,  which  is about
20\,Hz.  Elongations  in speckles do  occur, but these  are correlated
with  telescope elevation  and directed  along a  line leading  to the
zenith as DSSI  does not correct the atmospheric  dispersion.  This we
have  calibrated  out  by  deconvolving  our data  with  point  source
observations that have a dispersion  model built in (the elongation is
measured on  a bright star observed  at small zenith  distance $z$ and
scaled  as $\tan  z$  when deconvolving  the  object).  The  resulting
reconstructed images remain diffraction limited.

Table~\ref{tab:res}   lists  all   observations.   Its   first  column
identifies the  target by  the {\it Hipparcos}  number of  the primary
component, while  the next column shows which  component was observed,
the primary A  or the secondary B; AB stands  for both components. The
following  columns  contain the  Besselian  date  of the  observation,
filter ($R$ or $I$, see above), and the detection limits $\Delta m$ at
separations of 0\farcs15 and 1\arcsec. For resolved binaries, the last
three columns give the position angle $\theta$, separation $\rho$, and
magnitude  difference  $\Delta  m$,  while the  detection  limits  for
resolved binaries are $\Delta m$ relative to the primary component.

Four  binaries have  known  orbits.  Residuals  from our  measurements
(average in $R$ and $I$ filters) to those orbits are given in Table~2,
as a  consistency check  of the DSSI  calibration.  However,  they are
much larger than the DSSI errors of $\sim$1\,mas \citep{Horch2012} and
reflect the  quality of  the orbits.  The  4th column gives  the orbit
grade from \citep{VB6}.

\begin{deluxetable}{ r r r  c l   } 
\tabletypesize{\scriptsize}   
\tablenum{2} 
\tablecaption{Residuals to orbits
\label{tab:VB6} }                    
\tablewidth{0pt}     
\tablehead{ 
\colhead{HIP}  &
\colhead{(O$-$C)$_\theta$}  & 
\colhead{(O$-$C)$_\rho$}  & 
\colhead{Gr.} &  
\colhead{Reference}   \\
& \colhead{(\degr)} & \colhead{($''$)} &   &  
}
\startdata
9621 & $-$3.7 & $-$0.105 & 3 & \citet{Hei1996a} \\
82510 & $-$0.5 & 0.048    & 3 & \citet{Sca2003c} \\
95589 & 0.0    & 0.023    & 4 & \citet{Sca2015a} \\
97477 & $-$0.1 & 0.000    & 2 & \citet{WSI2006b} 
\enddata
\end{deluxetable}

\subsection{Comments on individual systems}

Comments on some observed objects  are given here. Orbital periods
of  newly resolved  pairs  are estimated  by  assuming that  projected
separation  equals  semimajor  axis.   Overall, we  resolved  six  new
subsystems in the secondary  components. One of them was independently
found  by other  team. The  two closest  subsystems are  tentative and
require confirming observations.

{\it  HIP 6268.}   The faint  secondary component  of this  binary was
pointed and  tentatively resolved  into a new  20-mas pair  Ba,Bb. The
2\farcs6  separation  between A  and  B  was  last measured  in  1967;
presently AB must  be wider than 2\farcs8, as it did  not fit into the
DSSI  field.  The  separation of  Ba,Bb implies  a period  of 1.3\,yr.
Given the small separation, the  relative photometry $\Delta R = 1.08$
mag and $\Delta I  = 1.42$ mag is not reliable and  it is premature to
conclude that Bb is bluer than  Ba.  The masses of Ba and Bb estimated
from their  luminosity are 0.6  and 0.4 ${\cal  M}_\odot$. Ironically,
the  main  component A  has  never  been  observed with  high  angular
resolution  and has  no RV  coverage, so  a similarly  close subsystem
Aa,Ab would remain undetected, if it existed.

{\it HIP 9583.}  Both A and B were observed separately, the separation
of AB  is 2\farcs3.  The data  quality is below  average because light
from the  other component  is getting into  the speckle frames  of the
observed component. This gives a  ``bright'' edge to these frames, and
it is not handled well by the reduction routines, causing lines in the
reconstructed images. But nonetheless,  the power spectra show no hint
of  any fringes,  so  there is  no  evidence for  binarity for  either
component.

{\it HIP 12764.}  The secondary component B of the recently discovered
5\farcs1 binary was observed  and resolved into a 0\farcs048 subsystem
Ba,Bb.   However, the  pair  is not  seen  in the  $R$-band, and  this
detection remains tentative.  Both A and B were also observed with the
speckle camera  at SOAR and unresolved \citep{SAM15}.   The period of
Ba,Bb  is estimated  at 2.6\,yr,  the masses  are 0.7  and  0.5 ${\cal
  M}_\odot$.    The main component A has an RV trend \citep{Bonavita2007}
also indicative of  a subsystem. Therefore, the DSSI  data reveal this
as a potential new 2+2 quadruple system.

{\it HIP 12925.} This is a quadruple system where the main star A is a
spectroscopic binary with yet unknown orbit and two visual companions.
The  components  A  and   D,  at  1\farcs9  separation,  were  pointed
separately and  found unresolved.   The pair AD  was also  observed at
Palomar  in  2013 \citep{Palomar},  with  no  new subsystems  detected
either.

{\it  HIP  74211.} The  B-component  of  STF~1916  was unresolved.  It
contains a spectroscopic subsystem (D.~Latham, 2012, private communication).

{\it HIP 74734.}  The secondary  component of the 5\farcs4 pair HO~547
is  resolved  into  a  new  subsystem Ba,Bb  with  a  relatively  wide
separation of 0\farcs28 but a high contrast.  It is barely seen in
  the $R$ channel, and its tentative measure in $R$ disagrees with the
  reliable  measure in  $I$.   We  estimate  the period  of Ba,Bb  as
50\,yr, the masses are 0.65 and 0.12 ${\cal M}_\odot$. This resolution
highlights the high dynamic range of DSSI.

{\it  HIP 77052.}  Both  components of  the 4\farcs4  nearby (parallax
68\,mas) pair  A~2230 ($\psi$~Ser, GJ~596.1) were  observed, and Ba,Bb
was  resolved at  0\farcs22.   This subsystem  has been  independently
discovered  by \citet{Rodriguez}.   We estimate  the masses  of nearly
equal stars Ba and Bb as 0.25 ${\cal M}_\odot$.  The position of Ba,Bb
in 2009.5 was (31\degr, 0\farcs22), similar to (27\degr, 0\farcs22) in
2015.5, so the pair may  have already completed a full revolution.  It
was  measured again  at  SOAR on  2016.141  at (21\fdg8,  0\farcs202),
demonstrating fast retrograde motion.  The three measures correspond to a
6.1-year orbit  with a semimajor axis  of 0\farcs19 and a  mass sum of
0.5 ${\cal  M}_\odot$, but it  is premature to present  this tentative
orbit  here.   Masses  of  Ba  and  Bb  will  eventually  be  measured
accurately  from  the  orbit  to  test  evolutionary  models  of  late
M-dwarfs,  while the main  component A,  a G5V  solar analogue  with a
constant RV, will serve  as a reference.  The 529-year  orbit of the outer
binary AB  recently determined by \citet{Gat2013b}  corresponds to the
mass  sum of  1.44 ${\cal  M}_\odot$, in  agreement with  the estimated
masses of all three stars.

{\it  HIP 81662.}  Both A  and  B at  11\farcs6 from  each other  were
pointed separately  and found unresolved. A  trend in the RV  of A has
been noted by \citet{Bonavita2007}. 


{\it HIP 85741.}  The secondary component of the  5\farcs3 pair HU~673
is resolved at 0\farcs2, with a  large $\Delta m$. The period of Ba,Bb
is $\sim$50yr, masses 0.55 and 0.14 ${\cal M}_\odot$.



{\it HIP 100970} hosts  a planet with 18-day orbit \citep{Fischer1999}
and has  a faint visual  companion B at  3\farcs5. We report  a secure
non-resolution of the main star, in agreement with other publications,
while B has not been observed. 

{\it  HIP 101345}  is  the pair  BU~668  with a  large contrast,  last
measured at  3\farcs2 in 1965  and now found  at 1\farcs2. It  has not
been resolved  by {\it Hipparcos}. 

{\it  HIP  110785}  is  a  triple system  consisting  of  the  913-day
spectroscopic  and astrometric binary  Aa,Ab \citep{Griffin2010}  in a
3\farcs6  binary BU~290 with  a poorly  constrained visual  orbit. The
estimated masses  of Aa and Ab  are 1.5 and 0.3  ${\cal M}_\odot$, the
semi-major  axis  of  Aa,Ab   is  59\,mas.   The  subsystem  Aa,Ab  is
unresolved here;  its highly inclined orbit means  that the separation
can often be smaller than the semimajor axis.

{\it HIP  111903} is  a 10\farcs9 pair  HU~3128 where  a spectroscopic
subsystem in  the primary component  A is suspected from  RV variation
\citep{CHIRON}. The secondary component B is unresolved here. 

{\it HIP  115417} has a newly resolved  secondary subsystem, discussed
in the next Section. There was some confusion about which component, A
or B, was pointed at  Gemini and resolved. Analysis of the astrometric
data indicates that the subsystem  belongs to B. This was confirmed in
2015   September  using   speckle   camera  at   the  SOAR   telescope
\citep{SAM15}.   The  SOAR  measure  of  Ba,Bb  (2015.7379:  181\fdg2,
0\farcs746, $\Delta m  = 1.97$ mag at 788\,nm)  agrees very well with
the Gemini measure (2015.5255: 180\fdg6, 0\farcs735, $\Delta I$ = 1.76
mag).


\section{The triple system HIP~115417}
\label{sec:orb}

HIP~115417 (HD~220334,  WDS 23228+2034) is a  nearby solar-type binary
with  the  following  properties:  spectral type  G2V,  HIP2  parallax
26.75$\pm$0.62  mas, proper motion  $(+314.6, -11.7$)\,mas~yr$^{-1}$
\citep{HIP2},   $V=6.62$  mag.    The  WDS   database   contains  262
measurements of AB. The  pair was frequently observed photographically
from 1942 to 1976 (mostly  at USNO), probably in search of astrometric
subsystems.

The visual binary AB was discovered by W.~Struve in 1829 at 5\farcs69
and 79\fdg5 (STF~3007). It has moved to 5\farcs85 and 92\fdg1 in 2007.
Estimated period  of this pair  is on the  order of 2\,kyr.  The Tycho
photometry gives $V$  magnitudes of 6.74 and 9.78  mag, $B$ magnitudes
of 7.42  and 10.98 mag for  A and B, respectively.  The distant visual
companion C, also discovered by Struve, is optical.

The motion  of AB observed for  nearly two centuries  shows a ``wavy''
character, most  obvious in  the position angles  and less  evident in
separations,  which are  measured  less accurately  than angles.   The
astrometric subsystem Ba,Bb responsible for this wave was first resolved by
us and confirmed at SOAR two months later. Archival observations of AB
allow us to estimate  orbital parameters of the subsystem.

\begin{figure}
\epsscale{1.1}
\plotone{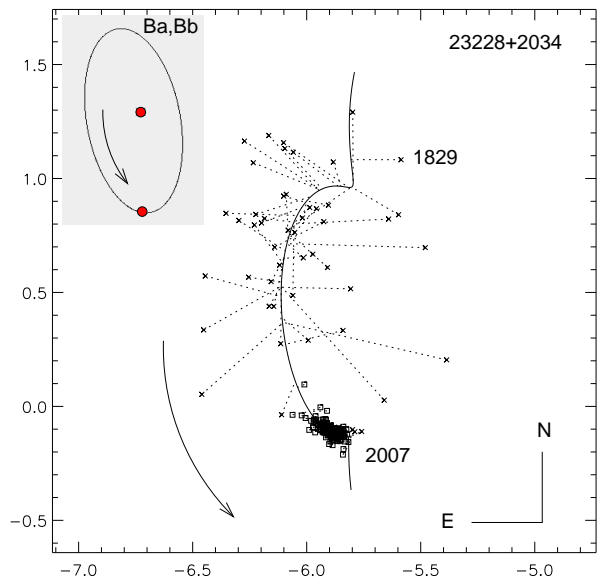}
\caption{Motion  of  the  component   B  relative  to  A,  located  at
  coordinate  origin,  in  the  binary system  HIP~115417  (STF~3007).
  Visual   observations   are   plotted   as   crosses,   photographic
  observations   as   squares.    The   line  shows   the   trajectory
  corresponding  to the  two orbits  proposed  here. The  scale is  in
  arcseconds.  The  gray insert  shows  the  orbit  of Ba,Bb  and  the
  observed position of this pair on different scale.
\label{fig:orb} }
\end{figure}

If the orbit  of the subsystem Ba,Bb is eccentric,  the center of mass
does not  coincide with  the center of  the orbital ellipse.   When we
subtract the smooth  motion of AB, this offset  is also subtracted and
the average residuals are zero; an astrometric orbit derived from
such residuals would have a small eccentricity, even if $e$ is in fact
substantial. This problem is solved by fitting the motion of AB and
the orbit of the subsystem Ba,Bb  simultaneously. We represent the motion of
AB  by a  circular  orbit and  adjust  its parameters  to obtain  the
expected mass  sum. Obviously,  the orbit of  AB is premature,  it is
derived here  only to  model its observed  segment, as needed  for the
analysis of the subsystem. The astrometric orbit of B (photo-center of
the  subsystem Ba,Bb)  is  described by  the  standard seven  Campbell
elements  ($P$--   period,  $T_0$  --  time  of   periastron,  $e$  --
eccentricity, $a$ -- astrometric  semimajor axis, $\Omega$ -- position
angle  of   node,  $\omega$  --   argument  of  periastron,   $i$  --
inclination).  An  additional parameter $F=A_2/a_2$, the  ratio of the
true  and astrometric  inner  axes, is  introduced,  and the  resolved
measures  of Ba,Bb  are added  to the  data set.   Thus, the  data are
modeled by 15 parameters.

The early visual measurements of AB are inaccurate and are compatible
with  a  wide range  of  orbital  parameters,  while the  photographic
measurements do  not cover the full  inner period. To  derive the most
likely  set of orbital  elements, we  estimate the  components' masses
from their luminosity using standard relations for main sequence stars
and choose the  elements that match those masses,  namely 1.2, 0.8, and
0.5 ${\cal M}_\odot$ for A, Ba, and Bb, respectively. This is achieved
by fixing  some elements in the least-squares  fitting.  By increasing
the inclination  of the  outer circular orbit,  we increase  the total
mass sum. The  inner mass sum is adjusted by  fixing the inner period.
The number of adjustable parameters is therefore reduced from 15 to 11.

In the least-squares adjustment of  11 free parameters, the weights are
inversely proportional  to the measurement errors,  which are assigned
subjectively  based  on  the  observing  method  and  then  corrected
iteratively to down-weight the outliers. The errors of most micrometer
measures are  taken as 0\farcs2,  the errors of  photographic measures
are 0\farcs02, in both  radial and tangential directions. The weighted
rms deviation from the  orbits is indeed 20\,mas. Figure~\ref{fig:orb}
shows  the measures  of AB  and the  trajectory corresponding  to the
elements listed in Table~3.

\begin{deluxetable}{ l c c   } 
\tabletypesize{\scriptsize}    
\tablenum{3}
\tablecaption{Orbits of HIP 115417
\label{tab:orb} }                    
\tablewidth{0pt}     
\tablehead{ 
\colhead{Element}  &
\colhead{AB}  & 
\colhead{Ba,Bb}   
}
\startdata
$P$ (yr) & 2161  & 117 (fixed) \\
$T_0$ (yr) & 1904.5 & 1935.1$\pm$5.5 \\ 
$e$       & 0 (fixed) & 0.204$\pm$0.04 \\
$a$ (\arcsec) & 6.053 & 0.215$\pm$0.011 \\
$\Omega$ (\degr) & 84.4   &  188.5$\pm$2.8 \\
$\omega$ (\degr) & 0 (fixed)   &   294$\pm$15 \\
$i$ (\degr)     & 58 (fixed)   &   65.1$\pm$2.5 \\
$F =A_2/a_2$  & \ldots &  3.21$\pm$0.17 
\enddata
\end{deluxetable}

If the light  of the component Bb is negligible,  the measures of
AB and A,Ba  are identical and the mass ratio  in the inner subsystem
is related to the ratio of the  axes, $q_2 = 1/(F -1)$ = 0.45 for
$F = 3.2$.   The estimated masses correspond to  $q_2 \approx 0.6$
or $F \approx 2.7$.  The light of Bb is thus slightly offsetting
the photo-center of B from the position of Ba, increasing $F$. 
This semi-qualitative analysis is all we can do at present, as the
data do not constrain both orbits sufficiently well to warrant a more
detailed investigation. The long inner period means that the situation
will not  improve soon.

\vspace{0.5cm}

\section{Summary}
\label{sec:concl}

We observed 25  secondary components of nearby binaries  with the DSSI
instrument  and  discovered  six  new  subsystems (one  of  those  was
independently found by others). Two  new pairs are very tight and have
short estimated  periods, illustrating the detection  power of speckle
interferometry  at a 8-m  telescope.  The  large fraction  of resolved
secondaries,     0.24,      supports     previous     results     {
  \citep{Tok2014,RAO,CHIRON}  and shows that secondary subsystems are
no less  frequent than subsystems  in the primaries.   Considering the
small number  of targets covered, it  makes little sense  to go beyond
this  qualitative statement  in  the statistical  analysis.  The  data
collected here, including  non-resolutions with deep detection limits,
will be used in the future to obtain a refined statistical analysis of
the  complete   67-pc  sample.   The  large   incidence  of  secondary
subsystems  suggests  that dynamical  interactions  in small  $N$-body
systems do not play a major role in the formation of multiple stars.

One  of the  newly resolved  secondary subsystems,  HIP  115417 Ba,Bb,
causes noticeable  deviation from  the slow motion  of the  outer pair
AB. We determined  a preliminary orbit of the  inner subsystem with a
period of 117 years using archival measurements from the WDS.

\acknowledgments 

We  thank all  observers  who participated  in  the DSSI  run. We  are
grateful to Gemini staff who aided in making the DSSI observations possible.
This work  used the  SIMBAD service operated  by Centre  des Donn\'ees
Stellaires  (Strasbourg, France),  bibliographic  references from  the
Astrophysics Data  System maintained  by SAO/NASA, and  the Washington
Double Star Catalog maintained at USNO.

{\it Facilities:}  \facility{Gemini-N}.


\LongTables

\begin{deluxetable}{ r c r c rr rrr  } 
\tabletypesize{\scriptsize}    
\tablenum{1}
\tablecaption{Summary of results
\label{tab:res} }                    
\tablewidth{0pt}     
\tablehead{ 
\colhead{HIP}  &
\colhead{Comp.}  & 
\colhead{Date}  & 
\colhead{Filt.}  & 
\colhead{$\Delta m(0.15)$} &
\colhead{$\Delta m(1)$} &
\colhead{$\theta$} & 
\colhead{$\rho$} & 
\colhead{$\Delta m$} \\
& & 
\colhead{(+2000) } & & 
\colhead{(mag)} &
\colhead{(mag)} &
\colhead{(\degr)} &
\colhead{(\arcsec)} &
\colhead{(mag)} 
}
  3203 & B & 15.5475 & R &    5.6 &    7.1 & \ldots & \ldots & \ldots\\
  3203 & B & 15.5475 & I &    5.3 &    7.8 & \ldots & \ldots & \ldots\\
  6268 & B & 15.5338 & R &    5.6 &    7.1 &    125.6 &    0.019 &     1.08\\
  6268 & B & 15.5338 & I &    5.6 &    7.1 &    124.8 &    0.022 &     1.42\\
  9583 & A & 15.5393 & R &    3.9 &    4.9 & \ldots & \ldots & \ldots\\
  9583 & A & 15.5393 & I &    4.3 &    5.3 & \ldots & \ldots & \ldots\\
  9583 & B & 15.5393 & R &    3.7 &    5.4 & \ldots & \ldots & \ldots\\
  9583 & B & 15.5393 & I &    3.2 &    5.4 & \ldots & \ldots & \ldots\\
  9621 & AB & 15.5338 & R &    5.5 &    7.5 &    343.2 &    1.398 &     2.30\\
  9621 & AB & 15.5338 & I &    5.6 &    7.2 &    343.2 &    1.397 &     2.18\\
 12764 & B & 15.5366 & R &    4.6 &    4.8 & \ldots & \ldots & \ldots\\
 12764 & B & 15.5366 & I &    4.8 &    5.2 &    237.2 &    0.048 &     1.37\\
 12925 & A & 15.5365 & R &    5.6 &    6.9 & \ldots & \ldots & \ldots\\
 12925 & A & 15.5365 & I &    5.3 &    7.5 & \ldots & \ldots & \ldots\\
 12925 & D & 15.5365 & R &    3.0 &    4.0 & \ldots & \ldots & \ldots\\
 12925 & D & 15.5365 & I &    4.9 &    5.9 & \ldots & \ldots & \ldots\\
 74211 & B & 15.5248 & R &    5.5 &    6.7 & \ldots & \ldots & \ldots\\
 74211 & B & 15.5248 & I &    5.5 &    6.7 & \ldots & \ldots & \ldots\\
 74734 & B & 15.5248 & R &    5.3 &    6.7 &    133.3 &    0.286 &     7.14\\
 74734 & B & 15.5248 & I &    5.4 &    7.2 &    130.2 &    0.282 &     4.39\\
 74771 & A & 15.5300 & R &    5.8 &    8.1 & \ldots & \ldots & \ldots\\
 74771 & A & 15.5300 & I &    5.9 &    7.1 & \ldots & \ldots & \ldots\\
 77052 & A & 15.5248 & R &    4.1 &    5.9 & \ldots & \ldots & \ldots\\
 77052 & A & 15.5248 & I &    4.1 &    5.6 & \ldots & \ldots & \ldots\\
 77052 & B & 15.5409 & R &    4.8 &    6.0 &     27.0 &    0.227 &     0.28\\
 77052 & B & 15.5409 & I &    4.8 &    5.6 &     26.9 &    0.226 &     0.35\\
 78217 & A & 15.5329 & R &    4.4 &    6.9 & \ldots & \ldots & \ldots\\
 78217 & A & 15.5329 & I &    5.4 &    7.2 & \ldots & \ldots & \ldots\\
 78217 & B & 15.5355 & R &    4.4 &    5.5 & \ldots & \ldots & \ldots\\
 78217 & B & 15.5355 & I &    4.9 &    6.4 & \ldots & \ldots & \ldots\\
 79169 & B & 15.5355 & R &    4.6 &    4.8 & \ldots & \ldots & \ldots\\
 79169 & B & 15.5355 & I &    5.1 &    6.0 & \ldots & \ldots & \ldots\\
 80467 & B & 15.5329 & R &    4.6 &    4.8 & \ldots & \ldots & \ldots\\
 80467 & B & 15.5329 & I &    4.7 &    5.0 & \ldots & \ldots & \ldots\\
 81662 & A & 15.5301 & R &    5.7 &    8.7 & \ldots & \ldots & \ldots\\
 81662 & A & 15.5301 & I &    5.6 &    8.3 & \ldots & \ldots & \ldots\\
 81662 & B & 15.5437 & R &    4.9 &    5.7 & \ldots & \ldots & \ldots\\
 81662 & B & 15.5437 & I &    4.6 &    6.7 & \ldots & \ldots & \ldots\\
 82510 & AB & 15.5330 & R &    5.5 &    6.9 &    104.6 &    1.443 &     1.95\\
 82510 & AB & 15.5330 & I &    5.5 &    6.6 &    104.9 &    1.438 &     1.88\\
 82886 & B & 15.5329 & R &    4.5 &    4.9 & \ldots & \ldots & \ldots\\
 82886 & B & 15.5329 & I &    4.9 &    5.5 & \ldots & \ldots & \ldots\\
 85307 & B & 15.5437 & R &    4.7 &    5.0 & \ldots & \ldots & \ldots\\
 85307 & B & 15.5437 & I &    5.1 &    6.0 & \ldots & \ldots & \ldots\\
 85741 & B & 15.5464 & R &    4.7 &    5.2 &    292.4 &    0.207 &     4.16\\
 85741 & B & 15.5464 & I &    5.2 &    6.1 &    293.8 &    0.210 &     3.24\\
 95589 & AB & 15.5250 & R &    3.6 &    5.0 &    289.3 &    2.195 &     1.02\\
 95589 & AB & 15.5250 & I &    4.2 &    5.2 &    289.7 &    2.186 &     0.82\\
 97477 & AB & 15.5251 & R &    4.8 &    6.9 &    110.2 &    0.488 &     0.50\\
 97477 & AB & 15.5251 & I &    4.8 &    6.9 &    110.5 &    0.487 &     0.66\\
 98677 & A & 15.5251 & R &    4.3 &    7.4 & \ldots & \ldots & \ldots\\
 98677 & A & 15.5251 & I &    4.5 &    7.6 & \ldots & \ldots & \ldots\\
 99367 & B & 15.5251 & R &    5.4 &    7.5 & \ldots & \ldots & \ldots\\
 99367 & B & 15.5251 & I &    5.2 &    7.5 & \ldots & \ldots & \ldots\\
100970 & A & 15.5251 & R &    5.3 &    8.3 & \ldots & \ldots & \ldots\\
100970 & A & 15.5251 & I &    5.2 &    8.0 & \ldots & \ldots & \ldots\\
101345 & AB & 15.5251 & R &    4.9 &    7.5 &     47.5 &    1.206 &     5.70\\
101345 & AB & 15.5251 & I &    4.8 &    7.9 &     47.8 &    1.182 &     6.11\\
104735 & A & 15.5281 & R &    5.7 &    9.3 & \ldots & \ldots & \ldots\\
104735 & A & 15.5281 & I &    5.9 &    8.7 & \ldots & \ldots & \ldots\\
107107 & A & 15.5281 & R &    5.5 &    8.9 & \ldots & \ldots & \ldots\\
107107 & A & 15.5281 & I &    5.9 &    8.3 & \ldots & \ldots & \ldots\\
110785 & A & 15.5253 & R &    4.2 &    6.8 & \ldots & \ldots & \ldots\\
110785 & A & 15.5253 & I &    4.7 &    5.9 & \ldots & \ldots & \ldots\\
111903 & B & 15.5281 & R &    4.9 &    5.5 & \ldots & \ldots & \ldots\\
111903 & B & 15.5281 & I &    5.7 &    7.1 & \ldots & \ldots & \ldots\\
112447 & A & 15.5255 & R &    4.9 &    6.2 & \ldots & \ldots & \ldots\\
112447 & A & 15.5255 & I &    4.8 &    6.7 & \ldots & \ldots & \ldots\\
113133 & B & 15.5281 & R &    5.7 &    7.8 & \ldots & \ldots & \ldots\\
113133 & B & 15.5281 & I &    5.5 &    7.3 & \ldots & \ldots & \ldots\\
115417 & B & 15.5255 & R &    4.4 &    6.5 &    180.7 &    0.736 &     2.39\\
115417 & B & 15.5255 & I &    4.6 &    6.3 &    180.6 &    0.735 &     1.76\\
116139 & B & 15.5281 & R &    4.6 &    5.1 & \ldots & \ldots & \ldots\\
116139 & B & 15.5281 & I &    5.6 &    6.9 & \ldots & \ldots & \ldots\\
116277 & A & 15.5255 & R &    4.3 &    4.7 & \ldots & \ldots & \ldots\\
116277 & A & 15.5255 & I &    3.9 &    4.5 & \ldots & \ldots & \ldots\\
117902 & A & 15.5282 & R &    5.6 &    8.8 & \ldots & \ldots & \ldots\\
117902 & A & 15.5282 & I &    5.8 &    7.6 & \ldots & \ldots & \ldots
\enddata
\end{deluxetable}

\end{document}